\documentclass[manuscript]{aastex}

\slugcomment{to be submitted to Astrophysical Journal}

\shorttitle{Astronomical ZeV Accelerator}
\shortauthors{Ebisuzaki and Tajima}

\begin{document}


\title{Astrophysical ZeV acceleration in the relativistic jet from an accreting supermassive blackhole Ver. 89.1
}


\author{Toshikazu Ebisuzaki\altaffilmark{1} and Toshiki Tajima\altaffilmark{2}}
\altaffiltext{1}{RIKEN, Wako, Saitama, Japan}
\altaffiltext{2}{University of California, Irvine, CA, 92679
}
\email{ebisu@postman.riken.jp}




\begin{abstract}
An accreting supermassive blackhole, the central engine of active galactic nucleus (AGN), is capable of exciting extreme amplitude Alfven waves whose wavelength (wave packet) size is characterized by its clumpiness. The pondermotive force and wakefield are driven by these Alfven waves propagating in the AGN (blazar) jet and accelerate protons/nuclei to extreme energies beyond Zetta-electron volt (ZeV$=10^{21}$ eV). Such acceleration is prompt, localized, and does not suffer from the multiple scattering/bending enveloped in the Fermi acceleration that causes excessive synchrotron radiation loss beyond $10^{19}$ eV. The production rate of ZeV cosmic rays is found to be consistent with the observed gamma-ray luminosity function of blazars and their time variabilities.
\end{abstract}





\section{Introduction}
The origin of ultra-high energy cosmic rays (UHECRs) with energies  $\> 10^{20}$ eV remains a puzzle of astrophysics. It is generally believed to be extragalactic (\cite{KO2011}, references therein). The production of UHECRs has been discussed mainly in the framework of the Fermi acceleration, in which charged particles gain energy through a numerous number of scatterings by the magnetic clouds. One of the necessary conditions of Fermi acceleration is the magnetic confinement: the Hillas criterion sets a constraint on the product of the magnetic field strength $B$ and extension $R$ of the candidate objects (Hillas criterion; \cite{Hil1984}): $W \le W_{\rm max}  \sim z (B/1{\rm \mu G})(R/{\rm 1kpc})  \, \rm EeV$, where $z$ is the charge of the particle. The possible candidate objects (but only marginally satisfy the Hillas criterion for  $10^{20}$ eV) are  neutron stars, active galactic nuclei (AGN), gamma-ray bursts (GRBs), and accretion shocks in the intergalactic space. However, the acceleration of $10^{20}$ eV particles even in those candidate objects is not easy for the Fermi mechanism because of 1) a large number of scatterings  necessary to reach highest energies, 2) energy losses through the synchrotron emission at the bending associated with scatterings, and 3) difficulty in the escape of particles which are once magnetically confined in the acceleration domain (\cite{KO2011}).

\cite{TD1979} first pointed out that intense electromagnetic fields create the plasma wakefield in which charged particles are accelerated by the Lorentz invariant pondermotive force in the longitudinal direction over a long distance. The wakefield acceleration has advantages over the Fermi mechanism in producing UHECRs for the following reasons: 
\begin{enumerate}
\item	The plasma wakefield provides an extremely high accelerating field.
\item	It does not require particle bending, which would cause severe synchrotron radiation losses in extreme energies.
\item The wakefield and particles move in the collinear direction at the same velocity, the speed of light, so that the acceleration has a built-in coherence, while the Fermi acceleration mechanism, based on multiple scatterings, is intrinsically incoherent and stochastic. 
\item	No escape problem exists. Particles can escape from the acceleration region since the wakefield naturally decays out. 
\item Whenever and wherever intense electromagnetic waves (with sufficiently high frequencies) are excited, such waves tend to exhibit coherent dynamics (see later for details).
\end{enumerate}
In fact the ``{\bf wakefield accelerator}''  concept is believed to be a promising candidate for a future-generation artificial accelerator (e.g. \cite{Lee2011}). It has been under intensive research (\cite{Esa2009}).

\cite{Tak2000} and \cite{Che2002} demonstrated that intense Alfven waves produced by a collision of neutron stars can create wakefields to accelerate charged particles beyond $10^{20}$ eV. Although such a neutron star collision is believed to be related to short gamma-ray bursts (\cite{Nak2007}), it is rather rare for two neutron stars to hit each other directly: It requires the same masses, otherwise the tidal field of the more massive star destroys the less massive one to form accretion disk. \cite{Cha2009} conducted one-dimensional numerical simulation showing that whistler waves emitted from an AGN produce wakefields to accelerate UHECR.

The accreting supermassive blackhole, the central engine of an AGN, is one of the candidates for wakefield acceleration. The accretion disk repeats transitions between highly magnetized (low-beta) state and weakly magnetized (high-beta) state (\cite{Shi1990}). In fact, \cite{ONe2011} have found that magnetic transitions with 10-20 orbital periods are predominant in the inner disk through their 3D simulation. Strong pulses of Alfven waves excited in the accretion disk at the transition can create intense wakefields in the relativistic jet launched from the innermost region of the accretion disk. Our analysis finds that these wakefields are natural to accelerate protons and nuclei up to extreme energies of ZeV ($10^{21}$ eV). 

\begin{figure}

\includegraphics[scale=0.7]{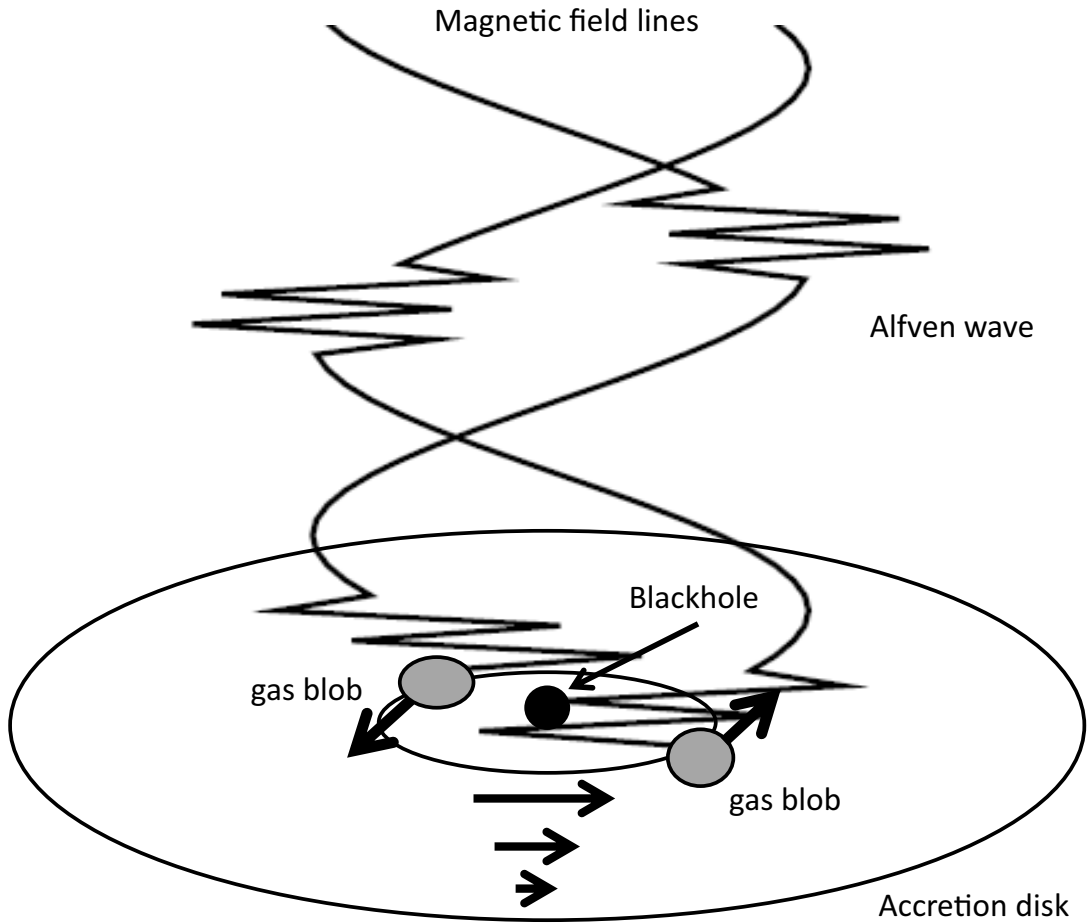} 

\caption{Schematic diagram of the production of intensive Alfven waves in an accretion disk. A gas blob formed near the inner edge of the accretion disk severely shakes the magnetic fields and excites relativistic Alfven waves, which propagate along the magnetic field line of the jet.}
\label{fig1}
\end{figure}

In the present paper we carry out a quantitative evaluation of the system of an accreting blackhole which consists of a blackhole itself, an accretion disk, and relativistic jets (figure \ref{fig1}), to lead to the generation of UHECRs beyond $10^{20}$ eV. The paper is organized as follows. We introduce our model for the intense wakefield generation that is not hampered by the Fermi mechanism limitations in Sec. 2 and find that the highest energy is achievable around an accreting blackhole (AGN) in Sec. 3. Astrophysical implications are discussed in Sec. 4.

\section{Intense wakefield mechanism}

An accretion disk is formed around a blackhole when gas accretes onto it. Since the angular velocity is higher in inner orbits, there arises a strong shear flow between gases circulating at different radii in the disk. Since the gas is almost fully ionized and Ohmic loss is negligible, magnetic fields are stretched and amplified by the shear motion. The resultant toroidal magnetic field acts as an enhanced friction between gases circulating in the different orbits and transfers the angler momentum outward, while gas is pushed inward because of the reaction of the momentum exchange.

The inner edge of the accretion disk is located around $R=3R_{\rm g}$, where 
\begin{equation}
R_{\rm g}=2GM/c^2=3.0 \times 10^{13} (m/10^8) \quad {\rm cm}
\end{equation}
is the gravitational radius of the blackhole. Here, $m$ is the mass of the blackhole in the unit of solar mass ($2.0\times 10^{33}$ g). An ergo-sphere appears just outside of causality horizon of the blackhole. The gas inside the ergo-sphere and the outside the horizon can extract the rotational energy from the blackhole, if they are magnetized. They drive relativistic jets in the two axial directions of the accretion disk (\cite{BZ1977}). The Lorentz factor ${\it \Gamma}$ of the bulk motion of the jet is observed as $10 \sim 30$ in the case of active galactic nuclei. 

According to \cite{Shi1990}, the accretion disk makes transition between two states: In the weakly magnetized state, magnetic fields are amplified by a strong shear flow, grow until at a certain point, and decay out, in other words, the disk makes transitions between these two states repeatedly. As a result, strong fluctuations are induced in the innermost region of the accretion region ($R<10R_{\rm g}$). The physical parameters in this innermost region ($R<10R_{\rm g}$) are estimated according to \cite{SS1973}: 
\begin{eqnarray} 
\varepsilon_{\rm D}=&6.6 \times 10^{6}& (m/10^8)^{-1} \quad {\rm erg \, cm^{-3}},\\ 
n_{\rm D}=&2.9 \times 10^{14}& (\dot{m}/0.1)^{-2} (m/10^8)^{-1} \quad{ \rm cm^{-3}},\\ 
Z_{\rm D}=&2.2 \times 10^{13}&  (\dot{m}/0.1) (m/10^8) \quad{\rm cm},\\ 
B_{\rm D}=&1.8 \times 10^3& (m/10^8)^{-1/2} \quad {\rm G},
\end{eqnarray}
where $\dot{m}$ is the accretion rate normalized to the critical accretion rate ($\dot{M}_c=L_{\rm Edd}/0.06c^2$; \cite{SS1973}). The viscosity parameter $\alpha$ is assumed to be 0.1 in the present paper. From the definition of $m$ and $\dot{m}$, the total luminosity of the accreting blackhole is given by 
\begin{equation}
L_{\rm tot}=1.3 \times 10^{45}(\dot{m}/0.1) (m/10^8)   \quad\rm erg \, s^{-1}.
\end{equation}
The wavelength $\lambda_{\rm A}$ of Alfven waves emitted from the accretion disk is calculated as (\cite{MT1995}):
\begin{equation}
 \lambda_{\rm A}=(V_{\rm AD}/C_{\rm sD})(\Omega /A)Z_{\rm D}=4B_{\rm D}Z_{\rm D}/3(4\pi \varepsilon_{\rm D})^{1/2}=5.8\times 10^{12} (\dot{m}/0.1)(m/10^8) \quad {\rm cm},
\end{equation}
 where $V_{\rm A}$ is the Alfven velocity in the accretion disk, which is caluculated as: 
\begin{equation}
V_{\rm AD}=B_{\rm D}/\sqrt{4\pi m_{\rm H} n_{\rm D}}=2.4\times 10^7(\dot{m}/0.1),
\end{equation}
and $C_{\rm sD}$ is the sound velocity in the accretion disk: 
\begin{equation}
C_{\rm sD}=\sqrt{\varepsilon_{\rm D}/m_{\rm H} n_{\rm SS} }, 
\end{equation}
where $m_{\rm H}$ is the proton mass. We assume magnetic field in the accretion disk as $B_{\rm SS}$ and the Keplarian rotation of gas inside the disk, i.e. $\Omega /A=4/3$.
The magnetic energy $E_B$ stored in the innermost region of the accretion disk ($R<10R_g$) is estimated as:
\begin{equation}
E_{\rm B}=(B_{\rm D}^2/4\pi) \pi(10R_{\rm g})^2 Z_{\rm D}=1.6\times 10^{48} (\dot{m}/0.1) (m/10^8)^2   \quad\rm erg.
\end{equation}
The Alfven waves excited in the accretion disk propagate along the global magnetic field of the jet. The normalized vector potential $a$ is the Lorentz-invariant strength parameter of the wave (\cite{Esa2009}) just above the disk is calculated as:
\begin{equation}
a_{\rm 0}=eE/m_{\rm e} \omega_{\rm A} c=2.3\times 10^{10}(\dot{m}/0.1)^{3/2}(m/10^8)^{1/2},
\end{equation}
where we used $E=(V_{\rm AD}/c)^{1/2}B_{\rm D}$ and $\omega_{\rm A}=2\pi V_{\rm AJ}/\lambda_{\rm A} \simeq 2\pi c/\lambda_{\rm A}$. The former is came of the conservation of Alfven energy flux,i.e., ${\it \Phi}_{\rm AJ}(=cE \times B/4\pi)= {\it \Phi}_{\rm AD}(=V_{\rm AD}B_{\rm D}^2/4\pi)$. We find that $a$ is much greater than unity for a large class of AGN disks and thus highly relativistic wakefield dynamics is anticipated (\cite{Ash1981}), where the Alfven flux inside of the jet assumed to be inversely proportional to $\pi b^2$ and $b$ to the square root of the distance $D$: $b=10R_{\rm g} (D/3R_{\rm g} )^{1/2}$. This scaling is consistent with the VLBI observation of the jet of M87, the closest AGN (\cite{Asa2012}). In such a case, $a$ of the wave propagating in the jet is calculated as:
\begin{equation}
a=a_0(D/3R_{\rm g})^{-1/2}, 
\end{equation}
where $D$ is the distance from the black hole along the jet.
The Lorentz factor $\gamma$ of the quivering motion of particles in the wave is of the order of $a$, i.e., $\gamma \sim a$. 

The kind of Alfven waves that are anticipated in the AGN jets have an extremely large value of $a$. It is a challenge for us to explore the physics of extremely relativistic wave dynamics (it is even beyond what has been called the "ultrarelativistic regime" (\cite{Mou2006}) and has been seldom considered. \cite{Ash1981} have considered such a regime in which protons begin to behave similarly to positrons. Bulanov et al. (2013) have analysed some aspect of the pondermotive acceleration. We shall take liberal indulgence into such works in the  past as our guidance in extending them for the present purpose.

The Alfven pulse generation, its collinear propagation feature, and its pondermotive acceleration are all under coherent dynamics. In other words, the phase between the specific wave and particles to be accelerated are tightly locked because the phase velocity of these waves (including the Alfven pulse in the jet under consideration) is very close to the speed of light. Further note that the wakefield in one dimension is robust because of the relativistic coherence (\cite{Taj2011}). The mechanism known as the dephasing (along with the pump depletion; \cite{TD1979}; \cite{Esa2009}) determines the maximum energy gain as well as the spectrum (\cite{Mim1991}; \cite{Che2002}).


\begin{figure}

\includegraphics[scale=0.6, angle=-90]{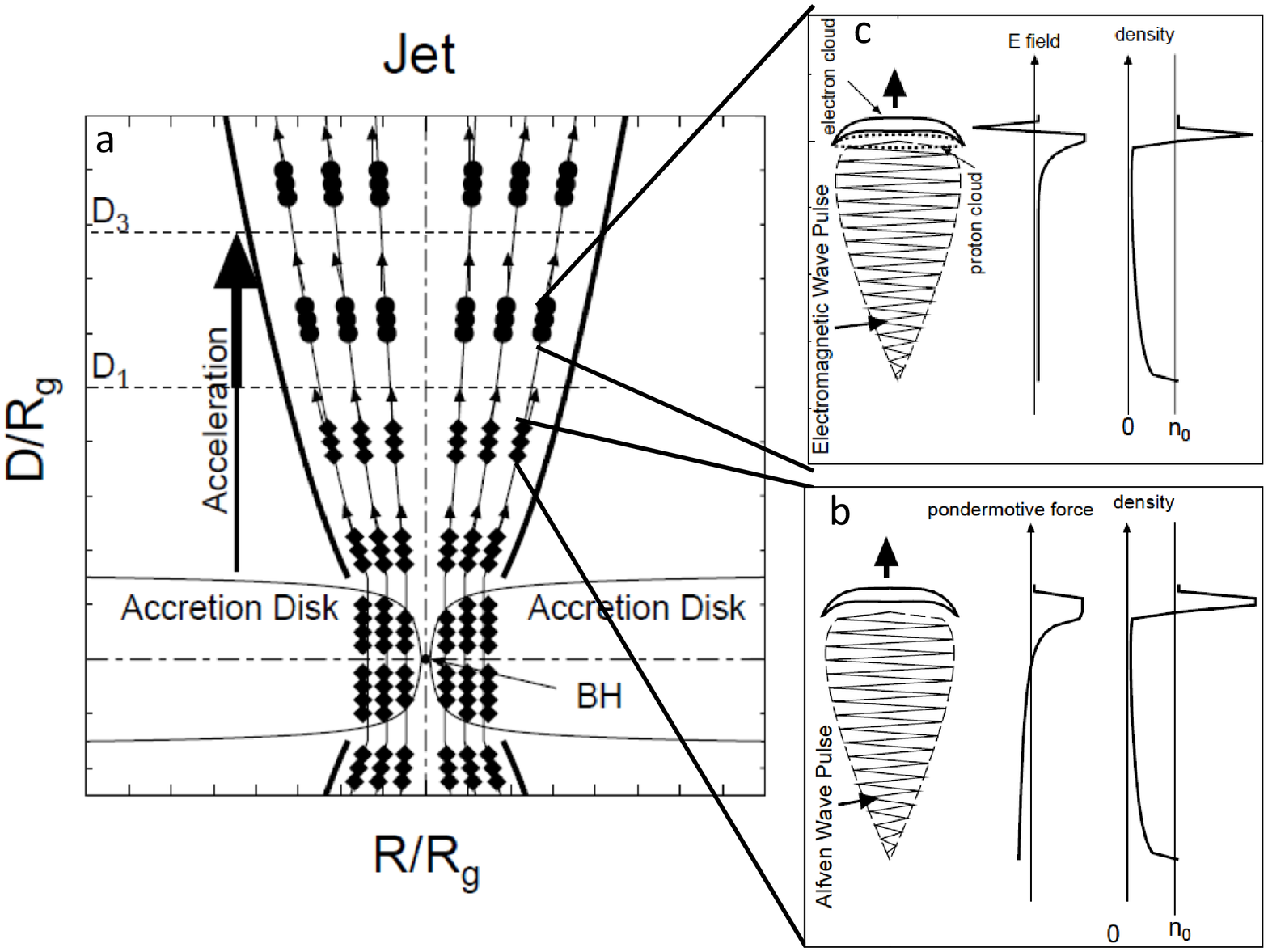}

\caption{a) Schematic cross section of an disk/jet system around an accreting black hole (BH). Alfven waves (diamonds) are excited in the accretion disk and propagates along the magnetic field (thin solid curves) in the relativistic jet (thick solid curves). b) In the pondermotive region ($\omega '_{\rm c}>\omega '_{\rm p}>\omega_{\rm A}$), the pondermotive force of the intense Alfven wave pulse produces a bubble and accelerate particles. c) The Alfven waves turn into electromagnetic waves (circles) as $\it \omega_{\rm A}$ approaches and exceeds $\it \omega'_{\rm p}$ and excites wakefields whose electric fields accelerate charged particles longitudinally along the jet. We anticipate that in the extremely large $a$, the domain of wakefield acceleration is dwarfed by that of pondermotive acceleration in the 1D situation. In 2-3D, wakefield acceleration takes a greater role than in 1D.}
\label{fig2}
\end{figure}


We focus on the wave modes propagating parallel to the jet magnetic field, since these modes are effective for the linear acceleration to highest energies.  The angular frequency of the Alfven wave is:
\begin{equation}
\omega_{\rm A}=2\pi V_{\rm AJ}/\lambda_{\rm A}\cong 2\pi c/\lambda_{\rm A}=3.2\times 10^{-2} (\dot{m}/0.1)^{-1} (m/10^8)^{-1} \quad\quad\rm Hz,
\end{equation}
where $V_{AJ}=B_{\rm J}/\sqrt{4\pi m_{\rm H} n_{\rm J}}$ is the Alfven velocity in the jet. If we assume the conservation of magnetic flux in the jet,
$B_{\rm J}$ in the jet is scaled as: 
\begin{equation}
B_{\rm J}=\phi B_{\rm D}(b/10R_{\rm g})^{-2}={\it \phi} B_{\rm D} (D/3R_{\rm g})^{-1},
\end{equation} 
where the plasma density $n_{\rm J}$ in the jet is calculated through the kinetic luminosity $L_{\rm J}$ of the jet, 
\begin{equation}
L_{\rm J}=n_{\rm J} m_H c^3 {\it \Gamma}^2 \pi b^2 =\xi L_{\rm tot}.
\end{equation}

\begin{equation}
n_{\rm J}=2.6\times 10^{3} (\dot{m}/0.1)(m/10^8)^{-1}(\xi/10^{-2})(\Gamma/20)^{-2}(D/3R_{\rm g})^{-1}\quad {\rm cm^{-3}}
\end{equation}
The effective plasma frequency $\omega '_{\rm p}$ is calculated as 
\begin{eqnarray} 
 \omega '_{\rm p}=&(4\pi n_{\rm J} e^2/m_{\rm e} \gamma\ {\it \Gamma}^3 )^{1/2}&\\
 =&2.1 \times 10^{-1} ({\it \Gamma}/20)^{-5/2}& (\xi/10^{-2})^{1/2}(\dot{m}/0.1)^{-1/4}(m/10^8)^{-3/4}(D/3R_{\rm g})^{-1/4} \quad \rm Hz,
\end{eqnarray} 
 On the other hand, the effective cyclotron frequency $\omega '_c$ is derived as 
\begin{equation}
\omega '_{\rm c} = eB_{\rm J}/m_{\rm e} c \gamma=2.3\times 10^{0}({\it \phi}/2.0)(\dot{m}/0.1)^{-3/2}(m/10^8)^{-1}(D/3R_{\rm g})^{-1/2}  \quad\rm Hz,	
\end{equation}

\begin{figure}
\includegraphics[scale=0.6,angle=-90]{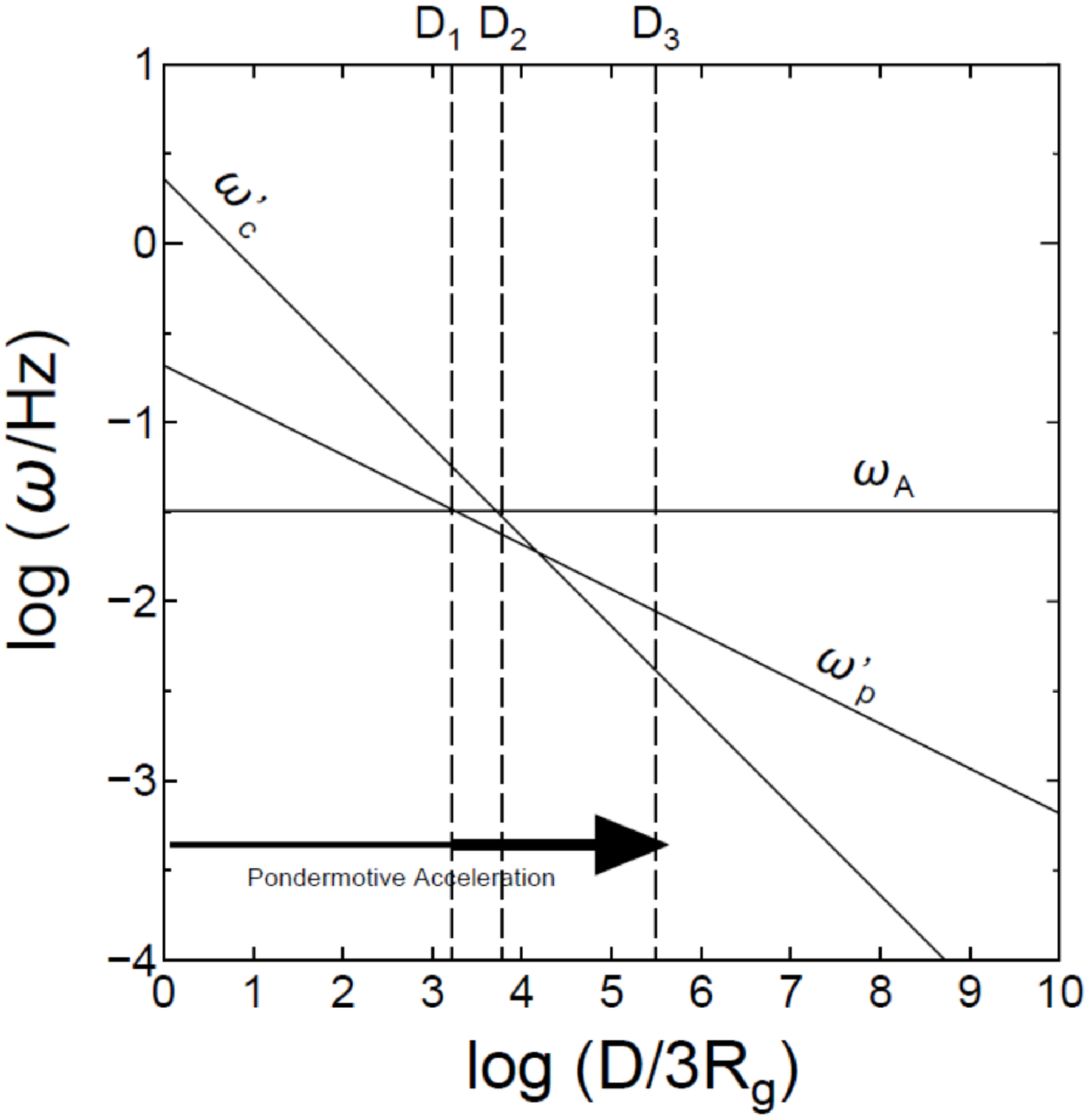}

\caption{Plasma frequency $\omega '_{\rm p}$ and cyclotron frequency $\omega '_{\rm c}$ are plotted against distance $D$ along the jet for the case of  $\dot{m}=0.1$, $\xi=10^{-2}$, ${\it \Gamma}=20$, and $m=10^8$. The pulse of Alfven wave with a frequency $\omega_{\rm A}$ is excited in the accretion disk ($D/R_{\rm g}=1$), propagating along the jet. Both $\omega '_{\rm p}$ and $\it \omega '_{\rm c}$  decrease as $D$ increases. The Alfven wave (whistler branch) turns itself into electromagnetic pulse around where $\it \omega '_{\rm p}=\omega_{\rm A}$ and drive wakefields to accelerate charged particles along the jet.  Further mode-conversions are possible beyond $\omega_{\rm A}>\omega '_{\rm c}$. In this figure, we show the dominance of the pondermotive acceleration in the extreme regime of $a \gg 1$ in 1-D. 
}
\label{fig3}
\end{figure}

As an Alfven wave pulse propagates along the jet, the density and magnetic fields decrease, and accordingly the ratios $\omega'_{\rm p}/\omega_{\rm A}$ and $\omega '_{\rm c}/\omega_{\rm A}$ plummet, as seen in figure \ref{fig3} (for the case of $\dot{m}=0.1, m=10^8, \gamma=20$, and $\xi=10^{-2}$).  As $\omega '_{\rm p}$ approaches $\omega_{\rm A}$, the whistler branch of the Alfven pulse turns into the electromagnetic wave (\cite{Cha2009}) and start to excite wakefields. The distance $D_1$ at which $\omega '_{\rm p}=\omega_{\rm A}$ is calculated as:
\begin{equation}
D_1/3R_g=1.7 \times 10^{3}({\it \Gamma}/20)^{-10}(\xi/10^{2})^2(\dot{m}/0.1)^{3}(m/10^8).
\end{equation}
On the other hand, the distance $D_2$ at which $\omega '_{\rm c}=\omega_{\rm A}$ is calculated as:
\begin{equation}
D_2/3R_g=5.1 \times 10^{3}(\dot{m}/0.1)^{-1}({\it \phi}/2.0)^2,
\end{equation}
independent to $\dot{m}$ nor $m$. As $d$ increases and $\omega '_{\rm c}$ approaches $\omega_{\rm A}$. In spite of the cyclotron resonance at $\omega '_{\rm c}$, most of the wave energy is likely to tunnel from the whistler branch to the upper branch beyond the right-hand cut-off frequency 
\begin{equation}
\omega_{\rm c}^{\rm rh}=[({\omega '_{\rm c}}^{2} + 4{\omega '_{\rm p}}^{2})^{1/2}+\omega '_c]/2,
\end{equation} 
which is located above the cyclotron resonance $\omega '_{\rm c}$ in the case of the cold and linear limit [\cite{Ich1973}]. In addition to the linear evanescent tunneling, the nonlinear nature of the EM waves (\cite{Ash1981}; $a \gg 1$) results in the resonance broadening and the nonlinear tunneling. 

\section{Highest energy cosmic rays}

The phase velocity of Alfven wave in the jet is close to the light velocity because of the small $n_{\rm J}$ compared to $n_{\rm SS}$. In such a case, the particles are accelerated by the pondermotive force parallel to the direction of the propagation of the wave. The maximum energy $W_{\rm PM}$ in the observer's frame of the particles  gained in the region is calculated as:

\begin{eqnarray}
W_{\rm max}=& z\int_{0}^{D_3} F_{\rm pm} dD&\\
=& 4.6\times 10^{19}& z({\it \Gamma}/20)(\dot{m}/0.1)^{1/2}(m/10^8)^{1/2}(D_3/3R_{\rm g})^{1/2}\quad{\rm eV}\\
=& 2.9\times 10^{22}& z({\it \Gamma}/20)(\dot{m}/0.1)^{4/3} (m/10^8)^{2/3} \quad {\rm eV},
\label{eqn:Wmax1}
\end{eqnarray}
(\cite{Ash1981}) where 
\begin{equation}
F_{\rm pm}={\it \Gamma}m_{\rm e} c a \omega_{\rm A}
\end{equation}
is the pondermotive force of the wave. The acceleration length is assumed to be:
\begin{equation}
Z_{\rm acc}=ca /\omega_{\rm A}.
\end{equation}
This is consistent with \cite{Ash1981}. Further, \cite{BM1990} obtained an exact nonlinear longitudinal plasma wave solution excited by a relativistic laser pulse, neglecting the quiver motion of protons.They found that the acceleration length is increased by a factor of $a$. This nature of acceleration lengthening can be expected to remain even in the case that proton quiver motion is not negligible, i.e., $a > 10^3$. They also found the plasma density is significantly reduced in the relativistic laser pulse because the plasma is evacuated by the strong pondermotive force. Equation \ref{eqn:Wmax1} holds as far as  $Z_{\rm acc}$
is greater than $D$. The distance $D_3$ is where the acceleration finishes, defined by the equation 
\begin{equation}
D_3=Z_{\rm pd}=ac/\omega_{\rm A}.
\end{equation}
We find that particles arrive at $D_1$ before $D_3$, in other words:  
\begin{equation}
D_3/3R_{\rm g}=3.9 \times 10^5(\dot{m}/0.1)^{5/3}(m/10^8)^{1/3}>D_1/3R_{\rm g}.
\end{equation} 

\begin{figure}
\includegraphics[scale=0.8]{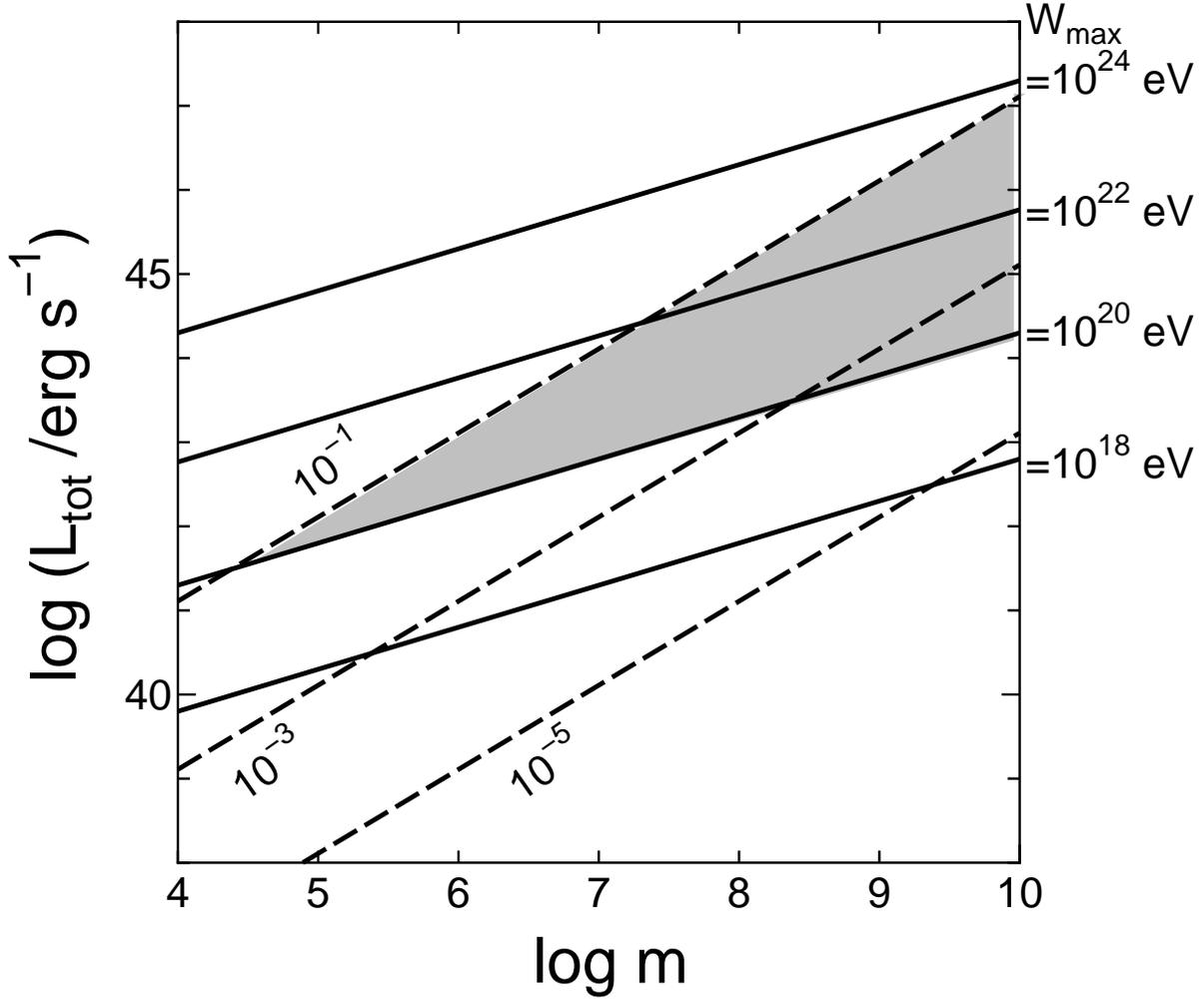}
\caption{The total luminosities of accreting blackholes are plotted against the blackhole mass (in the unit of solar mass) for various maximum attainable energy $W_{\rm max}$ (solid lines) for the case of ${\it \Gamma}=20$ and $\xi=10^{-2}$. Dashed lines represent the lines for $\dot{m}=10^{-5}, 10^{-3}$, and $10^{-1}$, respectively. The grey triangle represents the parameter sets which allow the acceleration of UHECRs ($ \geq 10^{20}$ eV).  We may set the upper limit of $\dot{m}$ to be around 0.1 for the pondermotive/wakefield acceleration may work, since the accretion disk becomes radiation dominant, as $\dot{m}$ approaches unity, so that an Alfven wave pulse becomes weaker than estimated in the present paper.
}
\label{fig4}

\end{figure}

The energy spectrum of the accelerated charged particles has the power-law with the index of -2 due to  the multiple dephasing occurrences when particles ride on and off different peaks of the pondermotive or wakefield hills when the waves contain multiple frequencies (but with again the same phase velocity $\sim c$; \cite{Che2002}), i.e.,
$f(W)=A(W/W_{\rm min})^{-2}$. As noted earlier, when the driving Alfven waves and their driven wakefields hold a broad band of frequencies, their phase velocities and group velocities, respectively, are again close to the speed of light, providing the basis for the robust accelerating structure. When wakefields have two or three dimensional features, the dephasing is prompted, leading to higher index of the spectrum (less than -2).
Let $\kappa$ be the energy conversion efficiency of the acceleration (including the mode convergence efficiency mentioned earlier), then $\kappa E_B =AW_{\rm min}^2 ln (W_{\rm max}/W_{\rm min})$,
i.e. \begin{equation}
A=1.6 \times 10^{33} \kappa \dot{m} m^2 [W_{\rm min}^2 ln (W_{\rm max}/W_{\rm min} )]^{-1}.
\end{equation}

The recurrence rate $\nu _{\rm A} $ of the Alfven pulse burst is evaluated as:
\begin{equation}
\nu _{\rm A}=\eta V_{\rm AD}/Z_{D} =1.0 \times 10^2 \eta m^{-1}   \quad\rm Hz,	
\end{equation}
where $\eta$ is the episode -dependent on the order of unity. This is consistent with the 3-dimensional simulations conducted by \cite{ONe2011}. They found magnetic fluctuations, called Long Period Quasi Periodic Oscillations (LPQPO) with the period 10-20 times the Kepler rotation period. The luminosity $L_{\rm UHECR}$ of ultra-high energy cosmic rays is: 
\begin{equation}
L_{\rm UHECR} \sim \kappa \zeta E_B \nu_{\rm A}=1.6\times 10^{33} (\kappa \zeta/0.01)\eta \dot{m} m  \quad \rm erg \, s^{-1},
\end{equation}
where $\zeta =ln(W_{\rm max}/10^{20} eV)/ln(W_{\rm max}/W_{\rm min} )$. 

The wakefields in the jets accelerate both ions and electrons and therefore the AGN jet is likely to be strong gamma-ray sources as well. Although the radiation loss of protons and nuclei is negligible as far as they are accelerated parallel to the magnetic field (\cite{Jac1963}), that of electrons is likely to be significant, when electrons encounter magnetic fluctuations. The gamma-ray luminosity is, therefore, found to be as:
\begin{equation}
L_{\gamma}\sim \kappa E_{\rm B} \nu_{\rm A}=1.6 \times 10^{34} (\kappa/0.1)\eta \dot{m} m  \quad \rm erg\, s^{-1}.
\end{equation}
We summarize the major features of pondermotive/wakefield acceleration in an accreting supermassive blackhole in Table \ref{table1}.

\clearpage

\begin{table}
\begin{center}
\caption{Major features of wakefield acceleration in an accreting supermassive blackhole.\label{table1}}
\begin{tabular}{clc}
\tableline\tableline
 & values& units \\
\tableline
$2\pi/\omega_{\rm A}$ & $2.0\times 10^2(\dot{m}/0.1)(m/10^8)$&s\\
$1/\nu_{\rm A}$ & $1.0\times 10^6\eta^{-1}(m/10^8)$&s\\
$D_3/c$ & $1.2\times 10^{9} (\dot{m}/0.1)^{5/3}(m/10^8)^{4/3}$&s\\
$W_{\rm max}$ & $2.9\times 10^{22} z({\it \Gamma}/20)(\dot{m}/0.1)^{4/3} (m/10^8)^{2/3}$& eV\\
\tableline
$L_{\rm tot} $& $1.2\times 10^{45}(\dot{m}/0.1)(m/10^8)$& ${\rm erg \,s^{-1}}$\\
$L_{\rm A}$& $1.2\times 10^{42}\eta(\dot{m}/0.1)(m/10^8)$&${\rm erg \,s^{-1}}$\\
$L_{\rm \gamma}$& $1.2\times 10^{41}(\eta\kappa/0.1)(\dot{m}/0.1)(m/10^8)$&${\rm erg \,s^{-1}}$\\
$L_{\rm UHECR}$& $1.2\times 10^{40}(\eta\kappa\zeta/10^{-2})(\dot{m}/0.1)(m/10^8)$&$ {\rm erg \,s^{-1}}$\\
$L_{\rm UHECR}/L_{\rm tot}$&$1.0\times 10^{-5}(\eta\kappa\zeta/10^{-2})$&-\\
$L_{\rm UHECR}/L_{\rm \gamma}$&$1.0\times 10^{-1}(\zeta/0.1)$&-\\
\tableline
\end{tabular}
\end{center}
$\xi=L_{\rm J}/L_{\rm tot}$, $\eta=\nu_{A}Z_{\rm D}/V_{\rm A}$, $\kappa=E_{\rm CR}/E_{\rm A}$, and $\zeta =ln(W_{\rm max}/(10^{20} eV))/ln(W_{\rm max}/W_{\rm min} )$.

\end{table}

\section{Astrophysical implications and blazar characteristics}
Radio galaxies belong to one category of AGN, which has radio lobes connected to the nucleus by relativistic jets. Their central engines are accreting supermassive ($m= 10^6-10^{10}$) blackholes. \cite{UP1991} pointed out that they are parent (or misaligned) populations of blazars, which show rapid time variations in many observational bands across radio to gamma rays (10 GeV) with distinct optical and radio polarizations because of their relativistic jet pointing almost toward us. The recent observation by the Fermi satellite reveals that many blazars emit strong gamma-rays in the GeV energy range (\cite{Har1999}; \cite{Ack2011}).  

We find that radio galaxies are most likely to be sources of UHECRs and their features fit well with the wakefield theory of acceleration. First, according to \cite{Aje2012} and \cite{Bro2012}, the local gamma-ray luminosity density of blazars is estimated as $10^{37-38} {\rm erg \,s^{-1}  (Mpc)^{-3}}$, taking into account the beaming effect of the relativistic jet. Assuming $L_{\rm UHECR}/L_\gamma \sim \zeta \sim 0.1$ (see table 1), our theoretical estimate of UHECR particle flux, averaged over the sky, becomes: 
\begin{equation}
\overline{{\it \Phi}_{\rm UHECR}}=7.6 \times 10^{-2}\, l_{\gamma 37}(\zeta/0.1)(\tau_8/1.5) \quad {\rm particles/(100\, km^2 \, yr\, sr)}. 
\label{eqn:flux_UHECR}
\end{equation}
Equation \ref{eqn:flux_UHECR} is consistent with observed flux of UHECR. Here, $l_{\gamma 37}$ is the local gamma-ray luminosity density of blazars in the unit of $10^{37} {\rm erg \, s^{-1} \, (Mpc)^{-3}}$ and $\tau_8$ is the life time of UHECR particles (in the unit of $10^8 {\rm yr}$), which is determined by GZK process: \cite{Gre1966} and \cite{ZK1966} predicted that cosmic-ray spectrum has a theoretical upper limit around $5 \times 10^{19}\,  {\rm eV}$, because of the opening of the channel to produce $\Delta ^+$ particles, which decay into pions ($\pi ^0$  and $\pi^{\pm} $) and further into photons, electrons, protons, neutrons, and neutrinos.  The flux of the cosmogenic neutrinos, produced by the GZK process, is as high as 
\begin{equation}
\overline{{\it \Phi}_{UHE\nu}}=5.4 \times 10^{-1}\, l_{\gamma 37}(\zeta/0.1)(\tau/100) \quad{\rm particles/(100\, km^2 \, yr\, sr)},
\end{equation}
assuming the conversion efficiency of UHECR to UHE$\nu$ to be $10\%$. This is consistent with the previous works for the case of $W_{\rm max}=10^{21.5}$ (e.g. \cite{Kot2010}). The recently observed PeV neutrinos with Ice Cube experiment (\cite{Arl2013}) is also consistent if we assume the power law spectrum of the index of -2.2 in the energy region from PeV to ZeV (\cite{Bar2013}). A next generation space borne detector of UHECR, like JEM-EUSO, which can achieve an integrated exposure of $10^6 \,{\rm km^2 \, str \, yr}$ (\cite{Tak2008}; \cite{Kaj2009}; \cite{Gor2011}), may detect this level of UHE$\nu$ flux (\cite{San2010}).

Second, blazars are also known for being highly variable at all wavelengths and all time scales. In the most extreme cases, the timescales of gamma-ray variability can be as short as a few minutes at very high energies ($\sim 100$ GeV; VHE). Such variability has been detected in several BL Lacertae objects (\cite{Arl2013}; \cite{Gai1996}; \cite{Alb2007}; \cite{Aha2007}; \cite{Ale2011}; \cite{Sai2013}). On the other hand, our wakefield acceleration mechanism predicts the rapid time variability with all the time scales from the Alfven frequency ($2\pi/\omega _{\rm A}\sim 100 \, {\rm s}$), through the repetition period of the pulses ($1/\nu_{\rm A}\sim$days), and to the propagation time in the jet ($D_2/c, 1\sim 10^2 \,{\rm years}$). This time variability is both for ion acceleration variability for UHECRs as well as electron variability as observed in gamma rays (electron energies are limited by the radiation energy loss by PeV (\cite{Den2012}). The finer structure of time variability is anticipated from our mechanism, as the magnetic structure may contain finer structure of braiding within the above quoted Alfven pulse. These observed blazer variabilities are the natural consequences deeply embedded in our model. Further, the coincidence of the pronounced luminosity peak and the reduced spectrum index observed Fermi satellite (\cite{Abd2010}) of BL Lacs has so far no known explanation offered but it is consistent with our theory (\cite{Aba2013}).

Third, the multiple epoch observation of VLBA provides a strong evidence that gamma-ray emission comes from parsec scale jet (\cite{Lis2009}; \cite{Mar2010}; \cite{Lyu2010}). Since the life time of high energy electrons is much shorter than the propagation time, they must be locally accelerated. This is consistent with the picture of the wakefield acceleration, since a swarm of the electrons is accelerated locally in the wakefields propagating in the jets. They are likely to emit a highly variable and polarized gamma-rays due to their high gamma-factor. 

\begin{figure}
\includegraphics[scale=0.6, angle=-90]{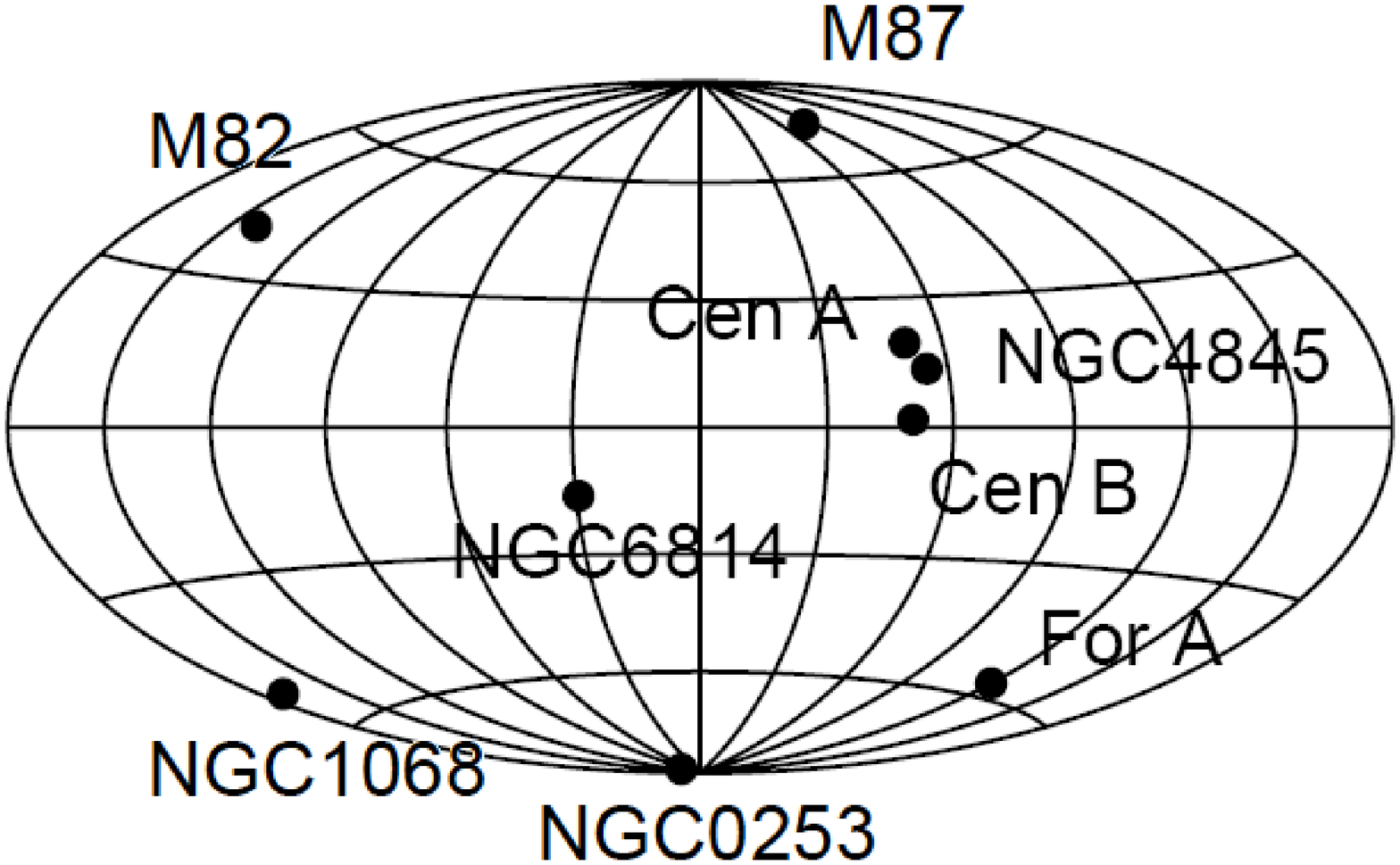} 
\caption{Distribution of the nine gamma-ray emitting AGNs (\cite{Ack2011}) in the sky
}
\label{fig5}
\end{figure}

Our calculation shows the UHECR flux of a gamma-ray emitting galaxy as:
\begin{equation}
{\it \Phi}_{\rm UHECR}=3.5 \times 10^{-3} (\zeta/0.1) ({\it \Phi}_{\gamma}/10^{-10}\, {\rm photons \, cm^{-2}\, s^{-1}}) (\overline{E_{\gamma}}/1\, {\rm GeV}) \\
\quad \rm particles/(100\, km^2 \, yr)^{-1},
\end{equation}
if the radiation pattern of UHECRs is the same as that of gamma-rays. Here, ${\it \Phi}_{\gamma}$ is the gamma-ray flux and $\overline{E_{\gamma}}$ is the average gamma-ray energy. We found nine gamma-ray emitting AGNs (\cite{Ack2011}) within the GZK horizon($\le 70$ Mpc; Table 2). The spectral indices are in the range of $-2\sim-2.8$, which are consistent with our theory. Figure 5 shows the distribution of the these gamma-ray emitting AGNs in the sky. This value is large enough to be identified as an individual source by a cluster of events with JEM-EUSO as well(\cite{Tak2008}; \cite{Kaj2009}; \cite{Gor2011}).

\begin{table}
\begin{center}
\caption{Nearby gamma-ray emitting AGNs detected by Fermi satellite (\cite{Ack2011})}\label{table2}
\begin{tabular}{crrlrc}
\tableline\tableline
Counterpart& LII &BII & Redshift& Flux (1GeV-100GeV) &Spectral Index\\
&&&&$10^{-10} \,{\rm erg \, cm^{-2}}$&\\
\tableline
NGC 0253 & 97.39&-87.97&0.001&$6.2\pm 1.2$&2.313\\
NGC 1068& 172.10&-51.04&0.00419&$5.1\pm 1.1$&2.146\\
For A&240.15&-56.70&0.005&$5.3\pm 1.2$&2.158\\
M82& 141.41&40.56&0.001236&$10.2\pm 1.3$&2.280\\
M87&283.78&74.48&0.0036&$17.3\pm 1.8$&2.174\\
Cen A Core&309.51&19.41&0.00183&$30.3\pm 2.4$&2.763\\
NGC 4945&305.27&13.33&0.002&$7.5\pm 1.7$&2.103\\
Cen B&209.72&1.72&0.012916&$18.6\pm 3.5$&2.325\\
NGC 6814&29.35&-16.02&0.0052&$6.8\pm 1.6$&2.544\\
\tableline
\end{tabular}
\end{center}
\end{table}

\section{Conclusions}
We have introduced the wakefield acceleration mechanism arising from the Alfvenic pulse incurred by an accretion disk around an supermassive blackhole, the central engine of AGN. This provides a natural account for UHECRs, and also the accompanying gamma-rays and their related observational characteristics, such as their luminosities, time variations, and structures. The severe physical constraints in the extreme ZeV energies by the Fermi acceleration have been lifted by the present mechanism. We have identified a number of areas of future research that needs further studies, including the bubble dynamics of super-intense Alfven pulses in 1-3 dimensions. We have already seen a number of emerging astrophysical phenomena that are not easy to explained by existing theories, but are in line with natural consequences of the present acceleration mechanism.

\acknowledgments

We would like to dedicated this paper to the late Professor Yoshiyuki Takahashi, whose encouragement on this work has been crucial. We appreciate the discussion with Profs. K. Abazajian, S.W. Barwick, S. Nagataki, A. Mizuta, and G. Yodh.

\clearpage

\end{document}